\documentclass[12pt]{article}
\usepackage{fullpage}
\usepackage{amsmath}
\usepackage{amssymb}
\usepackage[]{graphicx}
\graphicspath{{graphics/}} 

\usepackage{epstopdf}
\usepackage{multirow}
\usepackage{color}
\usepackage[linktocpage]{hyperref}
\usepackage{cite}
\usepackage{wrapfig}
\usepackage{subeqnarray}
\usepackage{lscape}
\usepackage[mathscr]{eucal}
\usepackage{dcolumn}
\newcolumntype{d}[1]{D{.}{\cdot}{#1} }

\usepackage{upgreek}

\def\blue{\textcolor{black}}

\def\c#1{\cite{#1}}

\def\le{\left(}
\def\ri{\right)}
\def\les{\left[}
\def\ris{\right]}
\def\lec{\left\{}
\def\ric{\right\}}

\def\##1{{\bf #1}}
\def\=#1{\underline{\underline #1}}
\def\4#1{\underline{\underline{\underline{\underline #1}}}}

\def\.{\mbox{ \tiny{$^\bullet$} }}

\def\eps{\epsilon}

\def\lambdao{\lambda_{\scriptscriptstyle 0}}

\def\ko{k_{\scriptscriptstyle 0}}

\def\uz{{\mathbf{u}}_z}

\begin{document}
\large
{\bf High-phase-speed Dyakonov--Tamm surface waves}

\normalsize

\vspace{10pt}
 
Tom G. Mackay$^{1,2,*}$  and
Akhlesh Lakhtakia$^{2}$

\vspace{10pt}

$^{1}$School of Mathematics and
   Maxwell Institute for Mathematical Sciences,
University of Edinburgh, Edinburgh EH9 3FD, UK;

 $^{2}$Department of Engineering Science and Mechanics, Pennsylvania State University,
	University Park, PA 16802--6812,
	USA

\vspace{10pt}

\begin{abstract}

Two numerical studies, one based on a canonical boundary-value problem and the other based on a reflection--transmission problem, revealed that the propagation of
high-phase-speed Dyakonov--Tamm (HPSDT) surface waves may be supported by the planar interface of a chiral sculptured thin film and a dissipative isotropic dielectric material.
Furthermore, many distinct HPSDT surface waves may propagate in a specific direction,  each with a phase speed  exceeding the phase speed of plane waves that can propagate in the isotropic  partnering material. Indeed, for the particular numerical example considered, the phase speed of an HPSDT surface wave could exceed the 
phase speed in bulk isotropic dielectric material by a factor of as much as 10.

\end{abstract}


{\noindent \footnotesize\textbf{*}Email:  \emph{T.Mackay@ed.ac.uk} }


\vspace{10mm}

\blue{
Any electromagnetic surface wave (ESW) is  bound to the interface of two dissimilar mediums \c{ESW_book}. Crucially,
at sufficiently large distances from the interface, the fields of an ESW decay in amplitude as distance from the interface increases.}
The planar interface of an isotropic material and an anisotropic material, with both materials being homogeneous, supports a particular type of \blue{ESW} known as a Dyakonov  wave \blue{\c{MSS,Dyakonov,Takayama1,Takayama2}}.
In a similar vein, the planar interface of a homogeneous material and a material that is periodically nonhomogeneous in the direction normal to the interface, with both materials being isotropic, 
 supports a particular type of \blue{ESW} known as a Tamm wave \cite{YYH,YYC,Shinn,Sinibaldi}.
A combination of these two cases gives rise to a Dyakonov--Tamm    wave, which is guided by the planar interface of a homogeneous isotropic dielectric material and a periodically nonhomogeneous anisotropic dielectric material \c{LP2007}.
Dyakonov--Tamm   waves are of particular interest because:
  (a) the range of their propagation directions is much larger than \blue{that} of Dyakonov waves \c{LP2007}; (b) air can be the isotropic partnering material in some instances \c{Faryad_JOSAB_2013}; and (c) multiple Dyakonov--Tamm 
waves can propagate in certain  directions \c{ESW_book,Faryad_JOSAB_2013}.
Owing to these characteristics, Dyakonov--Tamm  waves are promising candidates for applications such as optical sensing and energy harvesting. Their existence has been
experimentally verified \cite{PML2013,PMLHL2014}.

Usually, \blue{ESWs} propagate with phase speeds that are lower than  the phase speeds of plane waves propagating inside one of the partnering materials (if homogeneous). However, there are exceptions such as the high-phase-speed Tamm waves reported for the planar interface of two rugate filters
 \c{Maab}. Furthermore, very large phase speeds may be associated with optical Tamm states that may arise at the planar interface of two highly reflecting materials \c{Kavokin_PRB,Kavokin_APL}. These optical Tamm states should be distinguished from the surface states reported recently for the interface of a cholesteric liquid crystal and a uniaxial dielectric material \c{Timofeev}.

\begin{figure}
\begin{center}
\begin{tabular}{c}
\includegraphics[height=5.5cm]{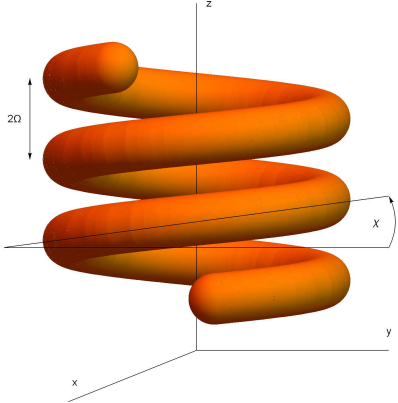}
\end{tabular}
\end{center}
\caption { \label{fig1}  \blue{A schematic representation of a CSTF's single helical column, with the structural period $2 \Omega$ and inclination angle $\chi$ indicated.}
} 
\end{figure}

Here we report on the  existence of high-phase-speed Dyakonov--Tamm (HPSDT) surface waves. For that purpose,
we consider  a Dyakonov--Tamm  wave guided by the planar interface of a structurally chiral material \cite{Collings,STFbook} and a dissipative isotropic  material, both dielectric and with permeability the same as of free space. Without loss of generality,
the planar interface is the plane $z=0$  and
 the Dyakonov--Tamm  wave is assumed to propagate parallel to the $x$ axis, with electric field phasor
\begin{equation}
\#E(\#r) = \#e(z) \exp \le i q x \ri,
\end{equation}
where \blue{$q\ne0$} is a complex--valued wavenumber and   
\begin{equation}
\#e(z) = e_x(z) \#u_x +
e_y(z) \#u_y +
e_z(z) \#u_z,
\end{equation}
with $\lec \#u_x,  \#u_y, \#u_z \ric$ being the triad of unit vectors aligned with the Cartesian axes.
The structurally chiral partner is a chiral sculptured thin film (CSTF)  characterized by the relative permittivity dyadic \c{STFbook}
\begin{equation}
\=\eps_{\,CSTF}(z) = \=S_{\,z} (z) \. \=S_{\,y} (\chi) \. \=\eps^{ref}_{\,CSTF} \. \=S^T_{\,y} (\chi)  \. \=S^T_{\,z} (z),
\end{equation}
where the reference relative permittivity dyadic 
\begin{equation}
\=\eps^{ref}_{\,CSTF}(z) = \eps_a \, \#u_z \, \#u_z + \eps_b \, \#u_x \, \#u_x + \eps_c \, \#u_y \, \#u_y
\end{equation}
specifies the local orthorhombic symmetry of the CSTF.
The rotation dyadics
\begin{equation}
\left.
\begin{array}{l}
\=S_{\,y} (\chi) = \le \#u_x \, \#u_x  +  \#u_z \, \#u_z \ri \cos \chi + 
\le \#u_z \, \#u_x  -  \#u_x \, \#u_z \ri \sin \chi + \#u_y \, \#u_y \vspace{8pt} 
\\
\displaystyle{ \=S_{\,z} (z) = \le \#u_x \, \#u_x  +  \#u_y \, \#u_y \ri \cos \le \frac{\pi z}{\Omega} + \psi \ri +
h  \le \#u_y \, \#u_x  -  \#u_x \, \#u_y \ri \sin \le \frac{\pi z}{\Omega} + \psi \ri  + \#u_z \, \#u_z}
\end{array}
\right\}
\end{equation}
are expressed in terms of the inclination angle $\chi\in(0^\circ,90^\circ]$, handedness parameter $h = \pm1$, structural period $2 \Omega$, and angular offset $\psi$ in the $xy$ plane. \blue{The CSTF is an array of  helical columns with each helical column being oriented parallel to the $z$ axis. A schematic illustration of a single helical column is provided in Fig.~\ref{fig1}.}

A variety of materials can be used to make CSTFs  \cite{STFbook} via physical vapor deposition
\cite{Mattox,HWbook}, with a range
of periods  $2\Omega$ and inclination angles $\chi$. Following Faryad \textit{et al.}
\cite{Faryad_JOSAB_2013}, for illustrative calculations let us choose   $\eps_a=3.965$, 
$\eps_b=4.917$, $\eps_c=4.361$, $\chi=67.53^\circ$, $h=1$, and
$\Omega = 160$~nm for the CSTF. Furthermore, let us set $n_s = 2.650 + 0.226 i$ as
the complex-valued refractive index of the
isotropic dielectric material.
\blue{This value of $n_s$ does not correspond precisely to any known natural material, as far as the authors are aware.  However, a material with such a  refractive index may be straightforwardly  conceptualized as a homogenized composite material arising from readily-available natural component materials\c{MAEH}.}
 The free-space wavelength $\lambdao=633$~nm for all data reported here.

The following two theoretical problems have to be  investigated: (a) a canonical boundary-value
problem which yields a dispersion relation for the wavenumber $q$ of the Dyakonov--Tamm 
wave, and (b) a more realistic reflection--transmission problem which yields absorptance data. 

\vspace{0.5cm}

\noindent \emph{(a) Canonical boundary-value problem.} Suppose that the CSTF occupies the half-space $z<0$ while the isotropic dielectric material occupies the complementary half-space $z>0$. Following a standard procedure \c{ESW_book},
Cartesian solutions 
to the frequency-domain Maxwell curl postulates are sought which decay as $z \to \pm \infty$ with the wavenumber $q$ being independent of $z$.
The
allowable values of $q$ arise as roots  of a dispersion relation  of the form ${\sf det}[{\=Y}(q)] =0$ where the matrix $[\=Y]$ is a function of $q$ and ${\sf det}$ denotes the determinant.

\begin{figure}
\begin{center}
\begin{tabular}{c}
\includegraphics[height=8.5cm]{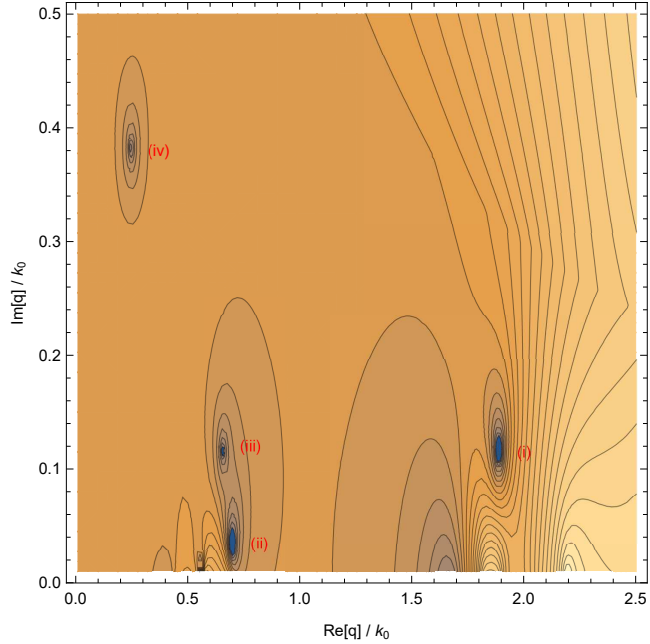}
\end{tabular}
\end{center}
\caption { \label{fig2} Scaled value of  $\ln\lec{\sf det}[{\underline{\underline Y}}(q)]\ric$ mapped against Re$\les q \ris/\ko$ and Im$\les q \ris/\ko$  for $\psi = 30^\circ$. The darkest regions contain the zeros of ${\sf det}[{\underline{\underline Y}}(q)]$. \blue{Solutions listed in  Table~\ref{tab1} are indicated.}
} 
\end{figure}

The scaled value of  $\ln\lec{\sf det}[{\=Y}(q)]\ric$ is mapped in the complex $q$ plane in Fig.~\ref{fig2} for $\psi = 30^\circ$.
The values of $q$  are normalized relative to the free-space wavenumber $\ko = 2 \pi/ \lambdao$. 
 Four roots of the dispersion relation may be deduced from
Fig.~\ref{fig2}. These are presented in Table~\ref{tab1} along with the respective
values of the phase speed $v_{DT} = \omega/\mbox{Re} \les q \ris$ relative to
 the phase speed in the bulk isotropic dielectric material, namely $v_s = \omega / \ko \, \mbox{Re}\les n_s \ris$, wherein $\omega$ is the angular frequency. For each of the four 
 allowable values of $q$ for surface-wave propagation, the phase speed $v_{DT}$ exceeds $v_s$. Therefore, each solution identifies an HPSDT surface wave.

\begin{table}[ht]
\caption{Solutions of the dispersion relation for surface-wave propagation} 
\label{tab1}
\begin{center}       
\begin{tabular}{|c|c|c|c|} 
\hline
\rule[-1ex]{0pt}{3.5ex}  Solution & $q/\ko$ & $v_{DT}/ v_s$  & $ \sin^{-1} \le \mbox{Re}\les q \ris/\ko \ri $\\
\hline\hline
\rule[-1ex]{0pt}{3.5ex}  (i) & $ 1.90 + \blue{0.12 i} $ & 1.39  & $-$\\
\hline
\rule[-1ex]{0pt}{3.5ex}  (ii) & $ 0.70 + 0.04i $ & 3.78   & $44.43^\circ$ \\
\hline
\rule[-1ex]{0pt}{3.5ex} (iii) & $ 0.65 + 0.12 i $ & 4.08   &  $40.54^\circ$ \\
\hline
\rule[-1ex]{0pt}{3.5ex}  (iv) & $ 0.25 + 0.38 i $ & 10.60  & $14.48^\circ$ \\
\hline
\hline 
\end{tabular}
\end{center}
\end{table} 

\begin{figure}
\begin{center}
\begin{tabular}{c}
\includegraphics[height=16.5cm]{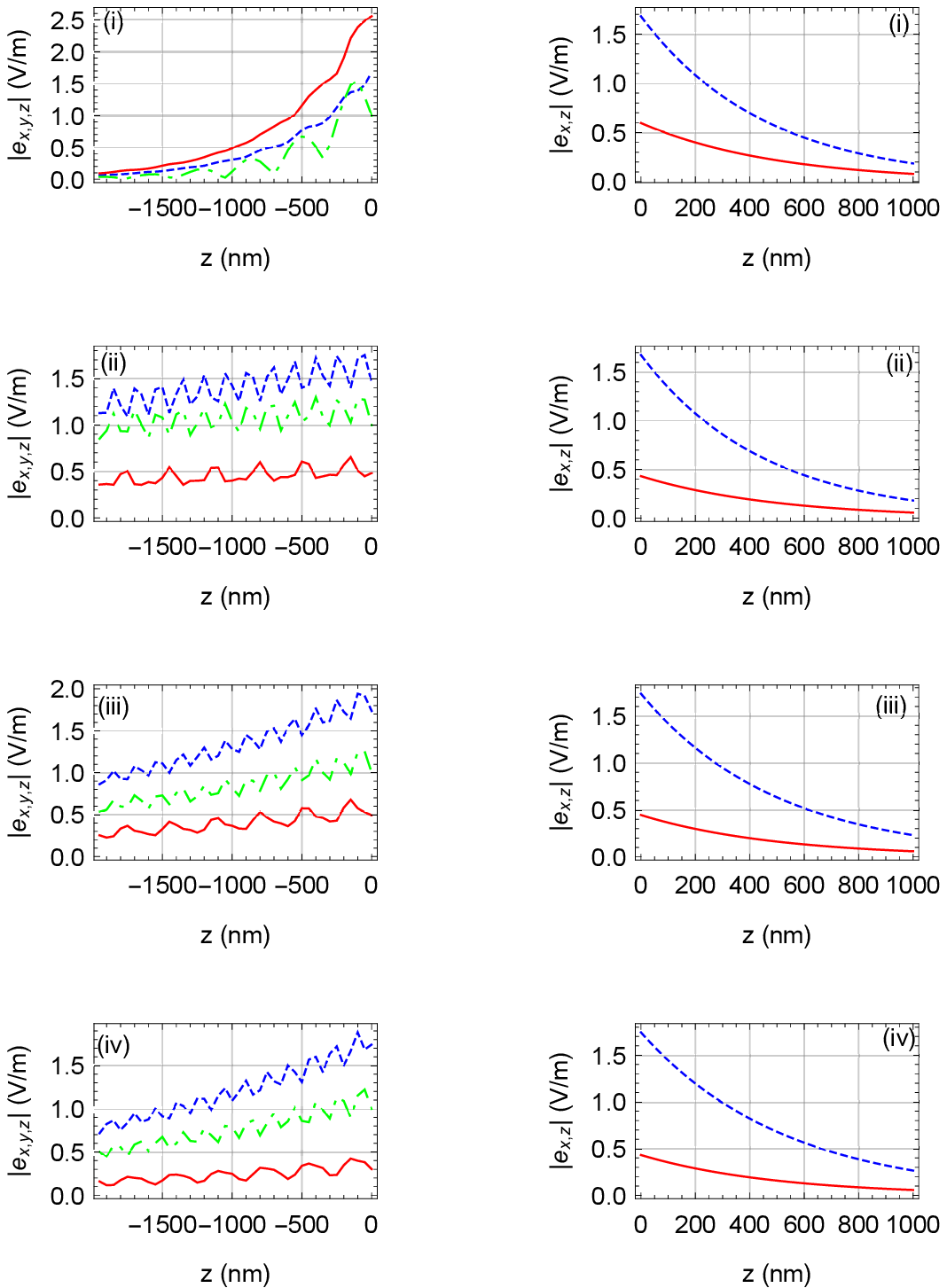}  
\end{tabular}
\end{center}
\caption 
{ \label{fig3}
Spatial profiles of normalized  magnitudes of the electric-field components in directions normal to the interface $z=0$ for 
   solutions (i)--(iv) given in Table~\ref{tab1}.  The normalization procedure is described in Ref. \citenum{ESW_book}.
   Key:  $|e_x|$  blue dashed curves,
 $|e_y|$ green broken-dashed curves,  $|e_z|$ red solid curves.} 
\end{figure} 

To explore this matter further, the spatial profiles of the absolute values of $\lec e_x, e_y, e_z \ric$ in directions normal to the interface are presented in Fig.~\ref{fig3} for each of the four solutions   (i)--(iv) given in Table~\ref{tab1}. 
These profiles reveal that the fields for each of the  solutions (i)--(iv) are localized to the vicinity of the interface $z=0$, with the degree of localization being greater on the side of the isotropic dielectric material (i.e., $z>0$) than on the CSTF side (i.e., $z<0$). Since the CSTF is nondissipative, the decaying profiles in the $z<0$ half-space provide further evidence in support of the excitation of HPSDT surface waves.

\blue{Even though ${\rm Re}[q]< \ko$ for solutions (ii)--(iv), these HPSDT surface waves cannot be classified as \textit{leaky surface waves}. This is because both partnering materials occupy half spaces so that the issue of leakage cannot arise.}

\vspace{0.5cm}

\noindent \emph{(b) Reflection--transmission  problem.}  Suppose that the CSTF occupies the finitely thick region $ - L_{CSTF}< z<0$ while the isotropic dielectric material occupies the
finitely region $L_s > z>0$. The half spaces $z < -L_{CSTF}$ and $z > L_s$
are  vacuous. An incident  plane wave propagates  in the half space $z < -L_{CSTF}$  towards the CSTF,
with its wavevector oriented at the acute angle $\theta_{inc}$ to  $\uz$. Therefore, a reflected plane wave propagates
 in the half space $z < -L_{CSTF}$ 
away from the CSTF, with its wavevector oriented at the acute angle $\theta_{inc}$ to  $-\uz$; and a transmitted plane wave propagates
 in the half space $z > L_{s}$ 
 away from the isotropic dielectric material, with its wavevector oriented at the acute angle $\theta_{inc}$ to  $\uz$. 
 
\begin{figure}[ht]
\begin{center}
\begin{tabular}{c}
\includegraphics[height=6.5cm]{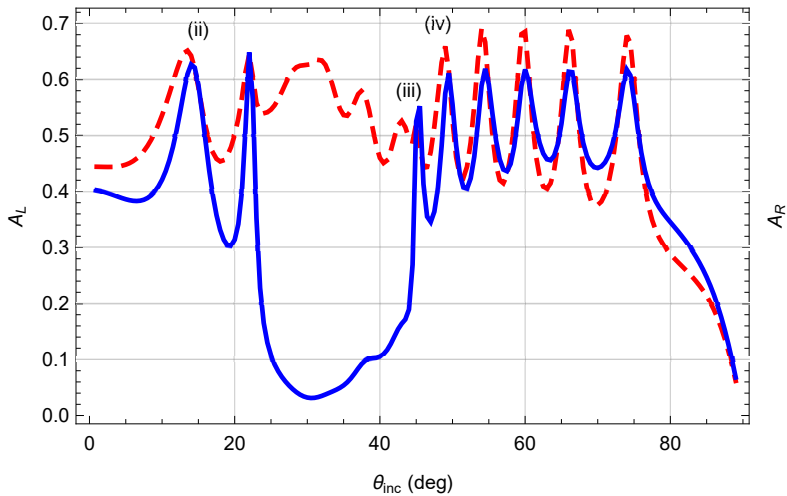}  
\end{tabular}
\end{center}
\caption 
{ \label{fig4} Absorptances $\mathcal{A}_L$ (red dashed curve) and $\mathcal{A}_R$ (blue solid curve) plotted against angle of incidence $\theta_{inc}$. Peaks corresponding to  solutions (ii)--(iv) in Table~\ref{tab1} are labeled.} 
\end{figure}

 In view of the structurally chiral morphology of the CSTF \cite{STFbook}, let   the incident plane wave be either left-circularly polarized (LCP) or 
 right-circularly polarized (RCP). Also, given the rates of decay of the electric-field components evident in Fig.~\ref{fig3}, let us choose $L_{CSTF} = 1.28 \,\mu$m (i.e., 40 CSTF periods) and $L_s = 200$ nm. The procedure for
 solving the reflection--transmission  problem and thereby
  computing the absorptances
for LCP incident light, namely $\mathcal{A}_L$, and for RCP incident light, namely $\mathcal{A}_R$, is comprehensively described elsewhere \c{STFbook}.
Graphs of $\mathcal{A}_{L}$ and $\mathcal{A}_{R}$ plotted versus $\theta_{inc}$ are provided in Fig.~\ref{fig4}. The $\theta_{inc}$-regime of relatively low values of
$\mathcal{A}_R$ as compared to $\mathcal{A}_L$, apparent in Fig.~\ref{fig4} extending 
over the range $22^\circ \lessapprox \theta_{inc} \lessapprox 46^\circ$, is a manifestation of the circular Bragg phenomenon by which means the CSTF discriminates between  incident LCP light and  incident RCP light \c{STFbook}.  

Values of the solution $q$  of the dispersion relation for the canonical boundary-value problem give an indication of the values of the angle of incidence $\theta_{inc}$ at which the excitation of HPSDT surface waves may be expected. Values of the angle of incidence are estimated as  $\sin^{-1} \le \mbox{Re}\les q \ris/\ko \ri $ and are listed in the rightmost column in Table~\ref{tab1}.
 Solution (i)  
 does not have a corresponding (real-valued) angle of incidence for the reflection--transmission problem, but solutions (ii)--(iv) do.

In Fig.~\ref{fig4} peaks in both  $\mathcal{A}_L$ and  $\mathcal{A}_R$ may be observed in the vicinity of the three  angles of incidence inferred from solutions (ii)--(iv). This observation provides further evidence in support of the existence of HPSDT  surface waves.

In conclusion, the 
two numerical studies presented here, one based on a canonical boundary-value problem and the other  on a reflection--transmission problem, provide strong evidence of the existence
high-phase-speed Dyakonov--Tamm surface waves that may be guided by the planar interface of a structurally chiral material and a dissipative isotropic dielectric material.
For the specific example investigated, the given propagation direction supported the propagation of four HPSDT surface waves with each 
 having a phase speed  exceeding the phase speed of plane waves in isotropic partnering material. Furthermore, in the case of the most extreme solution, 
  the phase speed of the Dyakonov--Tamm surface wave  exceeded the 
phase speed in  the bulk isotropic dielectric material by a factor of  10.

\blue{
Lastly,  the dissipative aspect of the isotropic dielectric material is not critical for the existence of HPSDT surface waves. Indeed, if the computations reported herein were repeated with the complex-valued refractive index $n_s$
replaced by the real-valued refractive index Re$\les n_s \ris$, then qualitatively similar results would be obtained, albeit the decay of the field profiles in the  $z>0$ half-space for the canonical boundary-value problem would be slower than the decay represented in  Fig.~\ref{fig3}.}\\

\noindent{\it Acknowledgment.}
AL is grateful to the Charles Godfrey
Binder Endowment at Penn State for ongoing support of his research.


\end{document}